\documentclass[fleqn,twoside]{article}
\usepackage{espcrc2}

\usepackage{graphicx}
\usepackage[figuresright]{rotating}

\newcommand{\AmS}{{\protect\the\textfont2
  A\kern-.1667em\lower.5ex\hbox{M}\kern-.125emS}}

\title{Atmospheric neutrino challenges}

\author{Todor Stanev\address{Bartol Research Institute, 
        University of Delaware, Newark, DE 19716, U.S.A. 
        }
        \thanks{This work is supported in part by U.S. 
        Department of energy contract DE-FG02 91ER. 
       }}
       
\begin{document}

\begin{abstract}
 We briefly review the improvements in the predictions of atmospheric 
 neutrino fluxes since the NOW2000 workshop. In spite of the great
 progress of the calculational technique the predictions are
 still not exact because of the uncertainties in the two major
 sets of input - cosmic ray flux and hadronic interactions on light
 nuclei. 
\vspace{1pc}
\end{abstract}

\maketitle

\section{INTRODUCTION}
 
 In 2004, after the experimental statistics on atmospheric neutrinos
 has become so good, one the major obstacles to the exact determination
 of the oscillation parameters is the uncertainty in the predictions
 of the atmospheric neutrino flux. The predictions are not bad, we 
 qualitatively understand well all features of the neutrino flux, but
 reaching the necessary 5\% or better level is still impossible.

 The basic features of the atmospheric neutrinos are 
 very well established. They follow from the neutrino production 
 processes and the development of the hadronic cascades in the
 atmosphere. The production mechanism is the decay chains of mesons created
 in these cascades. Positively charged pions, for example, decay into $\mu^+$
 and $\nu_\mu$. The muons subsequently decay into $\nu_e ,\; \bar{\nu}_\mu$
 and $e^+$. Since pion decays dominate the atmospheric neutrino production in
 the sub-GeV energy range, one can immediately predict the flavor ratio
 ${{\nu_\mu\; +\; \bar{\nu}_\mu} \over {\nu_e\; +\; \bar{\nu}_e}}$ = 2.

 The decay chain also determines the neutrino energy spectra.
 In the atmosphere mesons encounter the interaction--decay competition.
 Thus neutrinos from meson decay will have a spectrum one power
 of energy steeper than the primary cosmic ray spectrum. The muon
 daughter neutrinos will have a spectrum steeper by two powers
 of energy, because the muon  spectrum itself is steeper by $1/E$.
 Electron neutrinos thus have approximately $E^{-4.7}$ differential
 spectrum. Muon neutrino spectra are flatter. At low energy, however,
 the spectra are significantly flatter (parallel to the primary cosmic
 ray spectrum) as all mesons and muons decay. At high energy the
 neutrino spectrum is modified by the increasing kaon contribution,
 which asymptotically reaches 90\%. The contribution of charm and
 heavier flavors is still not essential.

 Neutrino energy spectra are a function of the zenith angle 
 of the atmospheric cascades. Mesons in inclined showers spend
 more time in tenuous atmosphere where they are more likely 
 to decay rather than interact. For this reason the spectra of
 highly inclined neutrinos are flatter than those of almost 
 vertical neutrinos.

 The general expectation is for up--down symmetric neutrino fluxes.
 The shorter distance to the atmosphere above a detector is 
 compensated by the smaller amount of atmosphere per unit solid 
 angle - both follow the $R^{2}$ law. 
 The symmetry is broken by the existence of geomagnetic field
 that prevents low energy cosmic rays from entering and interacting
 in the atmosphere in regions of low geomagnetic latitude. Such is the
 case in Japan where more low energy  atmospheric neutrinos enter
 the detector from below than from above. The symmetry should be
 restored at neutrino energies above 10 GeV (cosmic rays above 50 GeV)
 which are not affected by geomagnetic effects.
  
 There are two basic sets of inputs in a prediction of the atmospheric
 neutrino flux:\\
 -- Energy spectrum and composition of the cosmic ray flux. The 
 cosmic ray composition affects the ratio of neutrinos and 
 antineutrinos, thus the rate of neutrino events because of the
 different $\nu$ and $\bar{\nu}$ cross sections.\\
 -- Hadronic interactions on light nuclei (atmosphere) and particle
 production features in a wide energy range - from 1 to 10$^5$ GeV
 in the Lab.

 Neither of these sets of inputs has uncertainty of less than 10\%.
 Uncertainty estimates give higher values. 
 This is the basic reason for which the predictions of the atmospheric
 neutrino flux is a challenge. 

\section{GEOMETRY OF ATMOSPHERIC NEUTRINO PRODUCTION}

 Until recently, before NOW2000, all analyses of the atmospheric
 neutrino data were performed with the use of 1D calculations,
 such as Refs.~\cite{HKKM95,BGS89,AGLS}. These predictions are made
 with the assumption that all neutrinos follow the
 direction of the primary cosmic rays. Geomagnetic field was 
 only applied to the geomagnetic modification of the cosmic ray
 spectra at different locations as a function of the particle
 zenith angle. The situation now is quite different. There are
 more than seven independent calculations performed by 
 different~\cite{Bat00,HKKM01,Waltham,Wentz,Plyaskin,Batt03,Favier,Liu,BGLS04,HKKM04}
 groups.
\begin{figure}[thb]
\centerline{\includegraphics[width=60truemm]{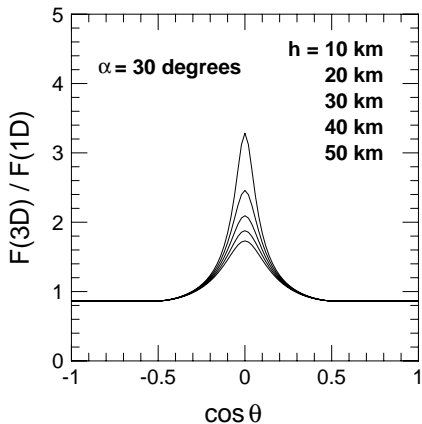}}
\vspace*{-20pt}
\caption{ Illustration of the 3D geometry for neutrino
 production angle 30$^\circ$ and different neutrino production
 heights.
\label{lipari1}
}
\vspace*{-10pt}
\end{figure}
 The geometry of the atmospheric neutrino production was first
 treated analytically and numerically by Lipari~\cite{Lipari00a,Lipari00b},
 who demonstrated that
 it is indeed very different from the 1D approximation. 
 In the realistic case when all secondary particles in the 
 atmospheric cascades are produced with transverse momenta   
 the neutrino angular distribution does not exactly follow
 that of the interacting cosmic rays. A vertical interacting
 cosmic ray generates neutrinos that are more inclined and 
 the effect is not compensated by cosmic rays of higher 
 inclination.
 The strength of the effects depends on
 the angle between the primary nucleon and the secondary 
 mesons and on the height of the {\em neutrino production}
 layer in the atmosphere. Figure~\ref{lipari1} illustrates 
 the effect for different production heights. There is a 
 peak of neutrinos around the horizon and a decrease
 of the neutrino flux at zenith angles less than 60$^\circ$.
 The total number of low energy neutrinos
 is somewhat higher than in the 1D approximation.
\begin{figure}[thb]
\centerline{\includegraphics[width=75truemm]{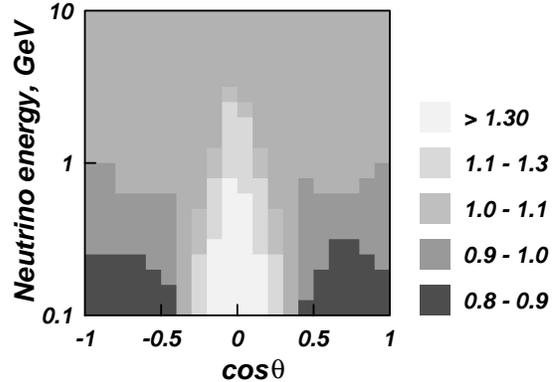}}
\vspace*{-20pt}
\caption{ Importance of the 3D effects at Kamioka in the \protect$E_\nu$
\protect$\cos {\theta_\nu}$ plane. 
\label{3dv1d}
}
\vspace*{-10pt}
\end{figure}
 It is obvious that the effect is strong at low neutrino energy - 
 30$^\circ$ production angle requires $p_\bot/p_\|$ ratio of 
 $\tan{{\rm 30^\circ}}$ = 0.58, i.e. relevant only for low energy
 interactions.
 At neutrino energies above 10 GeV, when neutrinos are produced
 under very small angle, the effect totally disappears. In practical 
 terms the 3D effects are not important above neutrino energy of 
 about 3 GeV. Figure~\ref{3dv1d} shows the importance of the 3D effects 
 in the energy/angle plane of the atmospheric neutrinos. The 
 biggest effect, an increase of the neutrino flux by more than 30\%
 is at the horizon for $E_\nu$ less than about 0.5 GeV. The 
 vertical neutrino fluxes below 0.3 GeV are decreased by up to 20\%.
 Fluxes for zenith angles around 60$^\circ$ are not changed by more
 than a couple of per cent. 

 At energies above 1 GeV there is only a slight (less than 10\%)
 increase of the fluxes below 3 GeV. At higher energy the effects
 totally disappear. 

 In spite of the greater sophistication of the modern 3D 
 calculations, most of the computer codes have to use 
 several simplifications. These include the treatment of
 the atmospheric density profile, which is often treated 
 as a single uniform profile (with exception of Ref.~\cite{Wentz}).
 Another is the altitude of the Earths surface, which is 
 of course different from sea level. The same is true for
 the altitude of the real detector for which the prediction
 is made - the SuperKamiokande detector is at about 3 km 
 higher altitude than SNO. A higher detector is obviously
 exposed to a slightly smaller flux of downgoing neutrinos.
 The most important  simplification if the size of the
 {\em neutrino detector} in the Monte Carlo simulation which
 is often bigger than  1000 km. Using very big detectors does
 not allow a correct  account for the local geomagnetic effects. 
 An interesting problem is the treatment of the low 
 Monte Carlo statistics of very inclined almost horizontal
 neutrinos, which can bring very large uncertainty in the 
 result.

 All these (and probably other) simplifications are necessary
 because of the very high ratio of the area of the Earths
 atmosphere to the area of even a huge detector. Their effects
 and not very well known and need a careful exploration. 
 They do not, however, contribute heavily to the uncertainties 
 in the neutrino prediction. These are dominated by the uncertainties
 in the basic inputs.

\section {PROBLEMS WITH THE BASIC INPUTS}

 Ten years ago the situation was worse: different measurements of the
 cosmic ray flux at about 10 GeV were different by about 50\%. 
 The situation has since improved, but not as much as
 we would like. 

\subsection{Cosmic Ray Spectrum}
 
 The agreement of the AMS~\cite{AMS} and BESS{\cite{BESS} data 
 on cosmic ray protons (Hydrogen nuclei) was considered (and it is)
 a break through.
 Protons are by far the most important cosmic ray nuclei in the
 energy range below 1000 GeV. They contribute 78\% of the all
 nucleon flux at 10 GeV, compared with the 15\% contribution
 of He and the total of 8\% contribution of all heavier
 nuclei. The better than 1\% agreement between the AMS and BESS
 results promised a serious improvement in our knowledge of the 
 cosmic ray flux. 

 The fluxes assigned to the He flux by these two experiments 
 do not agree that well, but the agreement is still better 
 than 10\%. The problem is that there are other measurements,
 particularly the CAPRICE results~\cite{CAPRICE98} that are 
 lower by about 20\% for both protons and He fluxes.
 Since the statistical and systematic errors given by the
 experimental groups are significantly lower than these
 differences one can not put together and fit all experimental
 data. The decision has then to be made: which experiments are
 good and which are not. I believe that scientists 
 outside the experimental group should not attempt to declare 
 an experiment more or less worthy. The only thing we can do is to 
 increase the size of the uncertainty on the cosmic ray flux.

 Including the contribution of heavier nuclei and extending
 the cosmic ray flux models to 10$^5$ GeV presents more 
 problems of similar character.
 Since it looks that  heavy nuclei have a flatter energy spectrum
 than protons it is quite possible than at energies above 10 TeV/nucleon
 the all nucleon spectrum could be dominated by the contribution
 of nuclei with $Z > 1$. This will change the cosmic ray composition - 
 the ratio of primary protons and nucleons interacting in the
 atmosphere. As an example we present in Fig.~\ref{alln} the
 all nucleon spectra derived in 1996 in Ref.~\cite{AGLS} and in 
 2001 in Ref.~\cite{HambCR}. The latter derivation does not include 
 CAPRICE data.
\begin{figure}[thb]
\vspace*{-10pt}
\centerline{\includegraphics[width=60truemm]{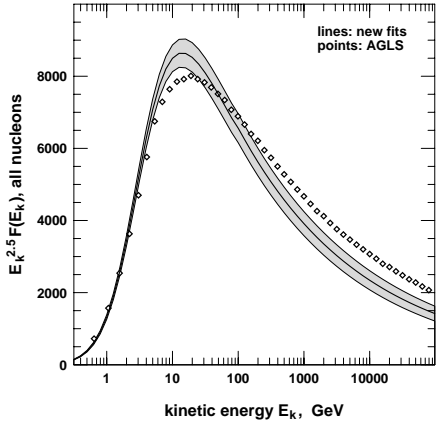}}
\vspace*{-20pt}
\caption{ All nucleon spectra derived before (AGLS) and after 
 the publication of the AMS and BESS data. The shaded area
 gives the derived uncertainties.
\label{alln}
}
\vspace*{-10pt}
\end{figure}
 At 10 GeV the newly derived flux is higher than the older
 parametrization by about 15\%. The two parametrizations cross
 over at about 50 GeV and the old parametrization is higher
 than the new one by more than 40\%  at 10$^5$ GeV.
 There are several reasons for the steeper all nucleon spectrum
 in the new fit. To start with, the highest energy several points
 in AMS and in BESS show a steep spectrum and with their lower
 errors dominate any fit. Secondly, the fluxes in the new 
 TeV measurement, RUNJOB~\cite{RUNJOB}, are lower than
 JACEE~\cite{JACEE}. That combination of high and low obviously
 requires a steeper spectrum. 

 It is not difficult to predict qualitatively what the effect
 of these two parameterizations is on the predicted neutrino  
 fluxes. The new sub-GeV neutrino fluxes will be higher 
 by 10-20\% while the high energy neutrinos, which are 
 responsible for the upward going muon events, will be lower. 

 One has to be very careful and inventive and use limits from 
 different experiments to constrain the flux models. As an
 example, the model of Ref.~\cite{HambCR} generates all
 particle flux at 10$^5$ GeV that is low compared to the
 estimates of the all particle flux from air shower measurements.
 
\subsection{Particle Production}

 The situation with the hadronic interaction models is not any better.
 There are a few measurements of the particle production on light nuclei,
 most of them on Be. Measurements were done when new neutrino beams
 were designed at accelerators. Most available data sets are from
 the 60's and the early 70's when the measurements were performed
 with single arm spectrometers, and correspondingly cover only a 
 part of the phase space. In the absence of applicable theory low energy
 hadronic interaction models are compiled to fit one or the other set
 of experimental data. Figure~\ref{pp}, compiled from Fig.~15 of
 Ref.~\cite{GH02}, gives an idea how different these
 models can be.
\begin{figure}[thb]
\vspace*{-10pt}
\centerline{\includegraphics[width=70truemm]{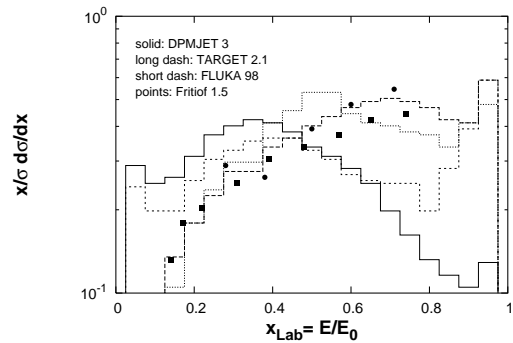}}
\vspace*{-20pt}
\caption{ Predictions of four interaction models are compared to 
 24 GeV p-Be\protect$\rightarrow$p data of Refs.~\protect\cite{Eichten,Allaby}.
 See Ref.~\protect\cite{GH02} for all references.
\label{pp}
}
\vspace*{-10pt}
\end{figure}
  DPMjet and FLUKA are qualitatively consistent (since FLUKA started
 as an extension of earlier DPMjet version). The homegrown model of
 the Bartol group, Target 2.1, is not very different from Fritiof 1.6,
 which is just an attempt to fit the data points shown in Fig.~\ref{pp}. 
 
 Fig.~\ref{pim} plots the $\pi^-$ production spectra in 24 GeV p-Be
 interactions predicted by Target 2.1 and FLUKA in comparison with 
 experimental data.
\begin{figure}[thb]
%%\vspace*{-10pt}
\centerline{\includegraphics[width=60truemm]{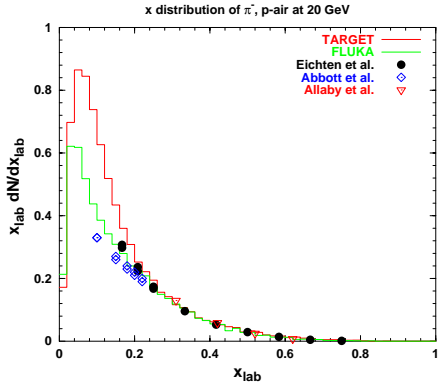}}
\vspace*{-20pt}
\caption{ Predictions of two interaction models are compared to 
 24 GeV p-Be\protect$\rightarrow\;\pi$ data of
 Refs.~\protect\cite{Eichten,Allaby,Abbott}.
\label{pim}
}
\vspace*{-10pt}
\end{figure}
 The differences between the two interaction models are significant.
 Visual inspection seems to show that Target 2.1 is a good description
 at high momenta, but may overestimate the pion production at low
 $x$ values. E-802 data favor the lower pion production model.
 
 At energies above 100 GeV the problems are different and 
 the main problem is the K/$\pi$ ratio as a function of energy.
 A large part of the problem is the associated $\Lambda K$ 
 production that generates hard positive kaons. The production cross 
 section is measured directly as well as by the $K^+/K^-$ ratio. 
 Models with large $\Lambda K$ production cross section predict
 high neutrino fluxes above about 100 GeV, where the kaon 
 contribution is already high~\cite{GH02}.
  
\section{UNCERTAINTIES}

 A recent study of the uncertainties in the prediction of the
 atmospheric neutrino fluxes from different hadronic interaction
 models was performed by Giles Barr 
 ({\em private communication}). The conclusions are that not
 all important atmospheric neutrino features are strongly
 affected by the model uncertainty. Flavor ratios
 ($\nu_e/\bar{\nu}_e, \; \nu_\mu/\bar{\nu}_\mu, \; 
 (\nu_\mu + \bar{\nu}_\mu)/(\nu_e/\bar{\nu}_e)$) are 
 stable to better than 1\% at $E_\nu$ less than 30 GeV. 
 At higher energies the uncertainties may reach 10\%.
 The up/down symmetry has a different behavior. Its
 uncertainty below 1 GeV if order 5\%, decreases to 1\% or less 
 between 1 and 10 GeV and then grows again to about 5\% at
 higher energy.

 The absolute normalization is a different
 story that depends on the cosmic ray flux as well as on the 
 interaction model. It is illustrated in Fig.~\ref{kaj21} 
 which compares the calculations of Refs.~\cite{Batt03,HKKM04,BGLS04}
 with the use of the same cosmic ray spectrum.
\begin{figure}[thb]
\vspace*{-10pt}
\centerline{\includegraphics[width=70truemm]{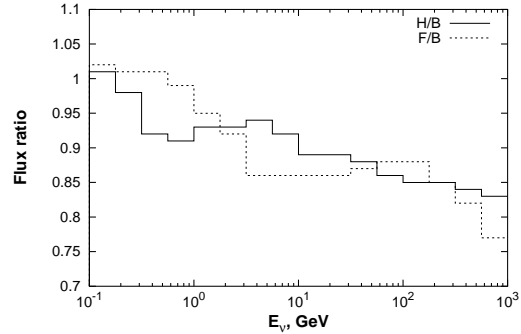}}
\vspace*{-20pt}
\caption{ Ratio of muon neutrino fluxes calculated by Refs.~\protect
\cite{HKKM04} (solid) and \protect\cite{Batt03} to the fluxes
 of Ref.~\protect\cite{BGLS04}.
\label{kaj21}
}
\vspace*{-10pt}
\end{figure}
 It is obvious that the Bartol calculation gives the highest 
 high energy neutrino flux. Up to energy of 10 GeV the Honda 
 et al. calculation~\cite{HKKM04} is on the 95\% level 
 and declines further at higher energy. The FLUKA calculation
 is lower than Bartol's already below 1 GeV and the difference increases 
 to more than 20\% when approaching energy of 1000 GeV. 

 Such differences affect mostly the analysis of the upward going 
 neutrino induced muon data, which depend on the neutrino flux
 of energy up to 10 TeV, where the differences could even be bigger.
 Figure~\ref{aglsvf} shows the angular distribution of upward going
 neutrino induced muons calculated with the fluxes of FLUKA~\cite{Batt03}
 and Agrawal et al.~\cite{AGLS}. The new Bartol fluxes are not used
 because currently they do not extend above 1000 GeV. 
\begin{figure}[thb]
\vspace*{-10pt}
\centerline{\includegraphics[width=70truemm]{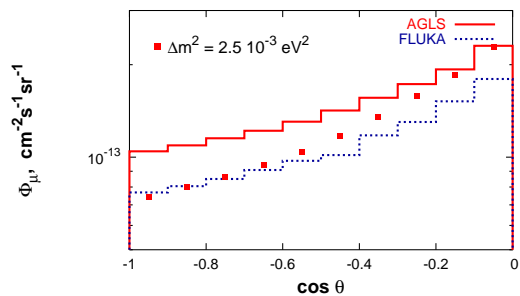}}
\vspace*{-20pt}
\caption{ Angular distribution of upward going neutrino induced muons
 at Kamiokande without oscillations - histograms. Points show Bartol's
 flux with oscillations (\protect$\Delta m^2$ = 0.0025 eV\protect$^2$).
\label{aglsvf}
}
\vspace*{-10pt}
\end{figure}
 The situation can be really confusing when the data (assuming it has
 the same shape as in our calculation) agrees with Bartol flux in
 horizontal direction and with FLUKA for vertical events without an
 account for oscillations. Although the Agrawal et al.~\cite{AGLS} flux,
 that was obtained with the old flux model in Fig.~\ref{alln},
 is higher than any other calculation the MACRO collaboration
 (see G.~Giacomelly in these Proceedings) concludes that the 
 new 3D calculations~\cite{Batt03,HKKM04} differ~\cite{MACRO04}
 from the global fit of their neutrino data.

\section{SUMMARY}

 The current situation with the predictions of the atmospheric
 neutrino fluxes is not ideal - the progress in theory seems to
 be behind that of experiments. On the other hand, there is 
 a big improvement in the calculational technique, which will
 eventually lead to significantly better results.

 To achieve significantly better predictions we need 
 much better input data on both the cosmic ray flux and 
 on the hadronic interactions on light nuclei. 

 The current program on studies of cosmic ray flux with balloon 
 instruments is strong, and we hope that these regular flights
 will continue during the next several years. We also expect the 
 satellite flights of the PAMELA and AMS experiments, that should
 be able to measure the cosmic ray flux even better. The keys are
 the achievement of understanding about the absolute normalization
 of different experiments and the extension of direct measurements 
 with the same instruments to energies above several TeV/nucleon.

 The hope for improvement of the hadronic interaction models is
 once again linked to the new neutrino beams that are prepared
 in both CERN and Fermilab. Parts of that preparation are the
 experiments HARP and P322 (using the NA49 detector) at CERN
 and MIPP (E907) at Fermilab. HARP 
 is in the process of data analysis, and P322 has had two runs at
 100 and 160 GeV. It would be a significant improvement over the past
 if the results of these experiments agree with each other. 

 The future is not bleak, but we do have a lot of work to do.

{\bf Acknowledgment} This talk was based on work performed 
 with G.D.~Barr, R.~Engel, T.K.~Gaisser, P.~Lipari and others.

\end{document}